\documentclass[aps,prb,amssymb,citeautoscript,showkeys,twocolumn]{revtex4-2}
\usepackage{graphicx}  
\usepackage{bm}       
\usepackage{amsmath}
\usepackage{subcaption}
\usepackage[labelsep=period]{caption}
\usepackage{hyperref}
\usepackage{dsfont} 
\usepackage{times}
\renewcommand{\vec}[1]{\bm{#1}} 
\hypersetup{colorlinks=true, linkcolor=blue, citecolor=blue, allcolors=blue}

\begin{document}

\title{Current- and field-driven domain wall dynamics and chirality switching in a planar helimagnet}

\author{Roman Teslia}
\author{Oleksiy Kolezhuk}
\affiliation{
 V.G. Baryakhtar Institute of Magnetism of the National Academy of Sciences of Ukraine, 03142 Kyiv, Ukraine
}%

\date{\today}

\begin{abstract}
We study the effect of electric current and magnetic field on the dynamics of chirality in a helimagnet with a strong easy-plane anisotropy. Using a continuum theory, derived for a quasi-one-dimensional frustrated ferromagnet close to the Lifshitz point, we show that a domain wall connecting domains with opposite chiralities can be driven by the current via the  dissipative (non-adiabatic) component of the spin-transfer torque.  Further, it is demonstrated that  the adiabatic part of the torque in the presence of a magnetic field breaks the symmetry of the effective magnetic potential energy with respect to the chirality. We show that the leading symmetry-breaking term arises in the third order in the magnetization gradient, and derive the conditions of the chirality switching. Our conclusions are supported by numerical spin-lattice simulations.
\end{abstract}

\maketitle

\section{Introduction} 
\label{sec:intro}

Frustrated magnetic systems have attracted an avid interest of researchers for several decades \cite{FrusMag2011}. 
Many of frustrated magnets belong to the class of so-called helimagnets, with the magnetization (or the Néel vector) forming a helical periodic structure in their ordered ground state. It has been demonstrated \cite{kishine2011tuning,wilson2013discrete} that the helimagnetic structure can be controlled by applying a magnetic field and the corresponding change of the magnetic state can be read by electric means.

Recent experiments show the possibility of chirality switching in conducting helimagnets by crossing the phase transition point with the application of electric current in the field-decreasing process \cite{jiang2020electric}, as well as near the paramagnetic state, assisted by the temperature \cite{masuda2024room}. Micromagnetic simulations \cite{ohe2021chirality} indicate that the controllability of the switching process significantly depends on the value of magnetic anisotropy. 

To our knowledge, however, the theory of chirality switching in helimagnets under the action of an electric current and a magnetic field, in particular the dynamics of current-driven chiral domain walls in these materials, remains largely unexplored. The main goal of this paper is to contribute to filling this gap. 

In this work, we theoretically investigate the motion of chiral domain walls under the action of an electric current in a quasi-one-dimensional centrosymmetric helimagnet, which has a doubly degenerate ground state (right- or left-handed helix), using a continuum description valid in the vicinity of the Lifshitz point where the period of the  helix is large compared to the lattice constant, and assuming a strong easy-plane anisotropy. 
We demonstrate the possibility of moving the chiral domain wall by the polarized electric current via the
dissipative (non-adiabatic) component of the spin-transfer torque, while the adiabatic part of the spin-transfer torque leads to the increase of the equilibrium pitch of the helix. 
Further, we show that the simultaneous application of magnetic field and polarized current gives rise to a term of the third order in the magnetization gradient in the effective potential energy, which breaks the symmetry with respect to the chirality and provides a mechanism for chirality switching. 
Those analytical predictions are supported by numerical simulations.

The structure of the paper is as follows: in Sec. \ref{sec:model} we introduce the model of a helimagnet, construct the continuum description incorporating the applied current and field, and discuss its range of applicability. 
In Sec. \ref{sec:helix} the analysis of effects of electric current on a single-domain state is given. In Sec. \ref{sec:ChirSwitch} we propose the chirality switching mechanism based on the chiral symmetry breaking caused by the simultaneous application of field and current. Sec. \ref{sec:dwdynamics} is dedicated to the dynamics of chiral domain walls. In Sec. \ref{sec:SLsimulations} we present the results of spin-lattice simulations and discuss their correspondence to the continuum model. Finally, Sec. \ref{sec:conclusion} contains a brief summary.

\section{Continuum model of a planar helimagnet} 
\label{sec:model}

Consider a quasi-one-dimensional (e.g., a nanowire) helimagnet oriented along the $x$ axis, defined by the Hamiltonian
\begin{equation}
\label{ham1}
\mathcal{H} = \sum_{\vec{r}} 
\Big\{
- J \vec{S}_{\vec{r}} \cdot \vec{S}_{ \vec{r}+a \hat{\vec{x}} }
- J'\vec{S}_{\vec{r}} \cdot \vec{S}_{\vec{r}+2a\hat{\vec{x}}} 
+ K(\vec{S}_{\vec{r}} \cdot \hat {x})^2
\Big\},
\end{equation}
where $\vec{S}_{\vec{r}}=S\vec{n}_{\vec{r}}$ are classical length-$S$ spin vectors at sites $\vec{r}$ of the lattice,  interacting via the ferromagnetic nearest-neighbor exchange $J>0$ and antiferromagnetic next-nearest-neighbor exchange $J'<0$, the easy-plane anisotropy strength is denoted by $K > 0$, and $a$ is the lattice constant  along the $x$ axis.  The hard  anisotropy axis  is assumed to be along the wire, so that the helical rotation occurs in the $(yz)$ easy plane.

For $|J'| > J/4$ the ground state is a simple helix, and it is convenient to introduce the helical angle between neighboring spins $\tilde \theta  = \arccos(- J/4J')$.  We assume that the helimagnet is close to the Lifshitz point $J=4|J'|$, then  $\tilde \theta\ll 1$  and  spin vectors are slowly varying at the scale of the lattice constant, so the continuum description can be used. 
Performing the expansion up to the fourth order in gradients, one obtains the following continuum Hamiltonian \cite{Hubert1974TheorieDD, melnichuk2002hubert,li2012vortex} for the unit vector field $\vec{n}(x)$:
\begin{equation}
\label{ham2}
\mathcal{H} = N_{\perp} \frac{ JS^2 a}{2} \int dx  \Big\{
\frac{a^2}{4} (\partial_x^2 \vec{n})^2 - \frac{\tilde{\theta}^2}{2} (\partial_x \vec{n})^2
+ \frac{2K}{Ja^2} n_x^2
\Big\},
\end{equation}
Here ${N_ \perp } = {S_ \perp }a/\mathcal{V}_0$ is the number of magnetic atoms  contained 
 in the cross-section area $S_ \perp$ of the wire, and $\mathcal{V}_0$ is the volume of magnetic unit cell.
In what follows, we set ${N_ \perp }$ to unity for the sake of simplicity, as it enters all expressions as the overall factor and does not affect the classical dynamics of the system. 

Let us briefly discuss the ground state of the system. 
The vector $\vec{n}$ can be parameterized by standard spherical angles: $\vec{n} = \sin \vartheta \cos \varphi  \cdot {\hat e_1} + \sin \vartheta \sin \varphi  \cdot {\hat e_2} + \cos \vartheta  \cdot {\hat e_3}$, where we choose ${\hat e_3} \equiv \hat x$. 
In the ground state all spins are confined to the easy plane, $\vartheta  = \pi /2$, and the Hamiltonian takes the following very simple form:
\begin{equation} 
\label{eq:HamGS}
    \mathcal{H}_\varphi=
    \frac{J S^2 a^2}{2}
    \int \frac{dx}{a}
    \Big\{
\frac{a^2}{4}
    \left[
    (\partial_x\varphi)^4 + ( \partial_x^2\varphi)^2
    \right]
    - \frac{\tilde{\theta}^2}{2} ( \partial_x\varphi)^2 
        \Big\}.
\end{equation}
Variation with respect to $\varphi (x)$ results in the equation 
\begin{equation} 
\label{eq:varPhiGS}
    2  ({\tilde \theta}/{a})^2 \partial_x^2\varphi  - 6  (\partial_x\varphi )^2 \partial_x^2 \varphi  + \partial_x^4\varphi  = 0,
\end{equation}
which admits the solution $\varphi  = q  x$ representing a uniform helix with an arbitrary pitch $q$. 
Substituting this solution into the Hamiltonian \eqref{eq:HamGS} and minimizing with respect to $q$ yields two degenerate ground states with $q =  \pm \tilde \theta /a$ corresponding to the right-handed (“$+$”) and left-handed (“$-$”) helicoids.
This twofold degeneracy allows for the formation of domain walls connecting the two ground states with opposite chiralities. 

\begin{figure}[tb]
\centering
\includegraphics[width=1.0\linewidth]{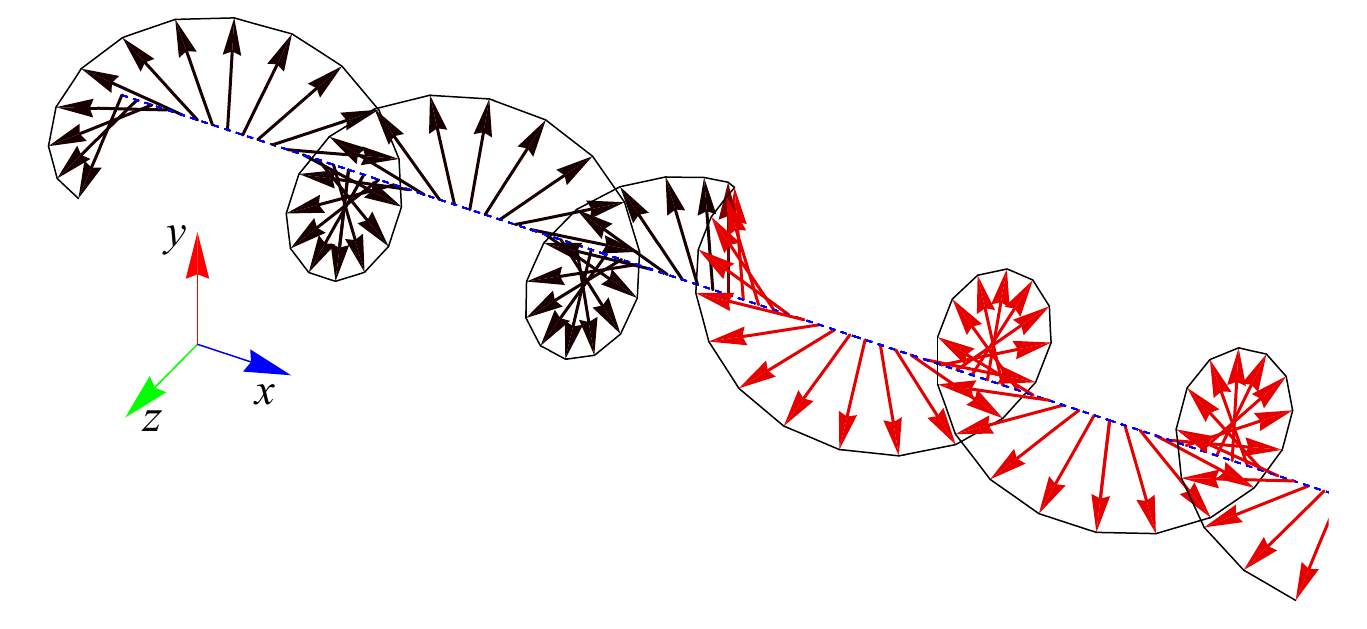}
\caption{Schematic depiction of an in-plane chiral (Hubert-type) domain wall connecting left-handed (on the left) and right-handed (on the right) helical domains. 
}
\label{fig:Hubert}
\end{figure}

The structure of such a chiral domain wall (DW) depends on the strength of the anisotropy. In the strong anisotropy limit $K\gg J$ it is favorable for the spins to stay in the easy plane.  The corresponding DW shape, sketched in  Fig. \ref{fig:Hubert},  is the exact solution of Eq. \eqref{eq:varPhiGS} first obtained by Hubert \cite{Hubert1974TheorieDD}:
\begin{equation} 
\label{eq:HubertDW}
    \varphi (x) = \pm \ln \cosh (\tilde{\theta} x/a).
\end{equation} 
The energy of this ``in-plane'' chiral DW (per unit  dimensionless area $N_\bot$) is easily obtained from Eq.  \eqref{eq:HamGS} as
\begin{equation} 
\label{eq:ChiralDWEnergy}
    E_{DW} =  JS^2\tilde{\theta}^3/3.
\end{equation}

In the limit of the weak anisotropy $K\ll J$ another type of the DW solution can be obtained, with the magnetization deviating from the easy plane. The pitch of the helix $q$ is determined solely by the exchange interaction, and for zero anisotropy the plane of rotation can be arbitrary.  The low-energy magnetization configurations can be written as $\vec{n}=1/\sqrt{2}(\vec{e}_a\cos qx +\vec{e}_b\sin qx)$, where $\vec{e}_{a,b}(x)$ is a pair of mutually orthogonal smoothly varying unit vector fields.   The DW solution connecting two helicoids with opposite chiralities can be constructed by changing the local plane of  rotation of spins in the helix, e.g., leaving $\vec{e}_b=\hat{\vec{y}}$ constant and letting $\vec{e}_a$ rotate from $\hat{\vec{z}}$ to $-\hat{\vec{z}}$  in the $(xz)$ plane. The energy of such ``out-of-plane'' DW can be obtained within the SO(3) continuum field description (see, e.g., \cite{DombreRead1989, AllenSenechal1995}) as
\begin{equation} 
\label{eq:CycloidalDWEnergy}
    E_{DW}^{\text{out-of-plane}} =  \sqrt{2JK}S^2\tilde{\theta}. 
\end{equation}
Comparing \eqref{eq:ChiralDWEnergy} and \eqref{eq:CycloidalDWEnergy} yields the condition for in-plane DWs to be energetically favorable
\begin{equation} 
\label{eq:favor-inplane-DW}
  \tilde{\theta}^2 \lesssim  K/J .
\end{equation}
In what follows, we assume that this condition is met, and leave the analysis of out-of-plane DWs for a separate study.

We are interested in the effects of the external magnetic field $\vec{H}$   and the electric current density $j^{(c)}$. The dynamics of the system can be described by the Langrangian  \cite{ShraimanSiggia88,Bazaliy98,kohno2007gauge,TATARA2008microscopic,kolezhuk2024current}
\begin{equation}
\label{eq:Lagr-cont}
L =
- \hbar S \int \frac{dx}{a}
\Big\{
\vec{A} \cdot (\partial_t \vec{n}+v \partial_x \vec{n}) -\frac{g\mu_B}{\hbar} (\vec{H}\cdot \vec{n})
\Big\} - \mathcal{H} .
\end{equation}
Here $\vec{A} = (\vec{n}_0\times \vec{n})/(1+\vec{n}_0\cdot\vec{n})$ is the vector potential of the Dirac monopole (the unit vector $\vec{n}_0$ defining the direction of the quantization axis $\vec{n}_0$ can be chosen arbitrarily), $g$ is the Landé factor and $\mu _B$ is the Bohr magneton. Further, 
\begin{equation}
\label{eq:st-velocity}
v =  j^{(s)} \mathcal{V}_0/(\hbar S) 
\end{equation}
is the so-called ``spin-transfer velocity'' \cite{kohno2007gauge} where ${j^{(s)}} = {j^{(c)}} P\hbar /(2e)$ is the spin current density, with  $P$ being the degree of spin-polarization of the current. The term with the spatial derivative in \eqref{eq:Lagr-cont} accounts for the adiabatic part of the spin-transfer torque, while the non-adiabatic part cannot be incorporated into the Lagrangian description and will be introduced later.
We will assume that the applied current is sufficiently weak compared to the largest scale of the problem set by the anisotropy constant:
\begin{equation}
\label{weak-curr-cond1}
v  \ll Ka/\hbar.
\end{equation}

Passing to spherical angles,  and assuming that the field is applied along the  hard axis $x$, and setting $\vec{n}_0=\hat{\vec{x}}$,  one can rewrite the Lagrangian in the following form:
\begin{eqnarray}
\label{eq:Lagr-cont1}
L &=&
\int \frac{dx}{a}
\Big\{S\big[ \hbar (\partial_t\varphi + v  \partial_x\varphi)+h\big]  \cos\vartheta
\Big\}- \mathcal{H} \nonumber\\
\mathcal{H} &=&
\frac{J S^2 a^2}{2}
    \int \frac{dx}{a} \sin^2\vartheta
    \Big\{
\frac{a^2}{4}
    \left[
    (\partial_x\varphi)^4 + (\partial_x^2 \varphi)^2
    \right] \\
   & -& \frac{\tilde{\theta}^2}{2} ( \partial_x\varphi)^2 - \frac{2K}{Ja^2} 
        \Big\}, \nonumber
\end{eqnarray}
where $h = g{\mu _B} H$. Here a few remarks are in order. 
In \eqref{eq:Lagr-cont1} we have dropped the full derivative terms contained in expressions of the type $\vec{A}\cdot \partial_{\mu}\vec{n}=(1 - \cos \vartheta ) \partial _{\mu}\varphi$ as they do not contribute to the equations of motion.  This is consistent with the fact that the adiabatic part of the current-induced spin transfer torque, taken alone (without the field), does not break the chiral symmetry. Further,  in \eqref{eq:Lagr-cont1} we have omitted terms proportional to $(\partial_x\vartheta)^2$, which  is justified for weak  currents as will be shortly shown below in a self-consistent manner.

Variation of the Lagrangian \eqref{eq:Lagr-cont1} over $\cos\vartheta$  allows us to obtain the angle of deviation from the  easy plane as follows:
\begin{equation} 
\label{eq:cosThetaFull}
    \cos \vartheta  = \frac{h+\hbar(\partial_t {\varphi} +v\partial_x \varphi)}{2KS} 
\Big\{ 1 - \frac{J\tilde{\theta}^2}{4K}(a\partial_x\varphi)^2
+\ldots\Big\} .
\end{equation}
Here we have made use of the relation $K/J \gg \tilde{\theta}^4$ that holds automatically as long as   \eqref{eq:favor-inplane-DW} is satisfied and $\tilde{\theta}\ll1$, so the terms in the curly braces are small corrections.  
Thus, in this approximation  $\vartheta$ is a ``slave'' field and the system behavior is fully determined by the dynamics of $\varphi$.
Space and time gradients of $\vartheta$ are proportional to second-order derivatives of $\varphi$,  so the inclusion of $(\partial_x\vartheta)^2$ terms in the Lagrangian \eqref{eq:Lagr-cont1} would lead to additional terms of the fourth order in gradients of $\varphi$.  They are small compared to the fourth-order terms already present in the Lagrangian if the  condition \eqref{weak-curr-cond1} is satisfied, which justifies the omission of $(\partial_x\vartheta)^2$ terms.
 We also neglect renormalization of $(\partial_x\varphi)^4$ terms coming from the magnetic field, as it carries the extra factor $J\tilde{\theta}^4/K\ll1$. 
Similarly, additional terms proportional to $(\partial_x\partial_t\varphi)^2$ can be safely neglected when considering the low-frequency dynamics.

Substituting  \eqref{eq:cosThetaFull} into the Lagrangian \eqref{eq:Lagr-cont1}, one obtains the effective Lagrangian depending only on  the ``master'' field $\varphi$: 
\begin{equation}
\label{eq:LQuadratic1}
\tilde{L} =
\int \frac{dx}{a}
\Big\{
\frac{\hbar^2 (\partial_t\varphi + v\partial_x\varphi)^2}{4K}  + \tilde{h} \frac{\hbar v S}{a}  F[\partial_x\varphi] - (1-\tilde{h}^2)\mathcal{H}_\varphi 
\Big\},
\end{equation}
where we have denoted
\begin{equation}
\label{eq:htilde}
\tilde{h}={h}/{(2KS)},
\end{equation}
and $F[\partial_x\varphi]$ is a functional that is odd under $\partial_x\varphi \mapsto -\partial_x\varphi$ and describes breaking of the chiral symmetry  which occurs when the external field and current are applied simultaneously,
\begin{equation}
\label{eq:oddTerms}
F[\partial_x\varphi]=  a\partial_x \varphi +\frac{\zeta}{3} (a\partial_x \varphi)^3 +\ldots
\end{equation}
However, the first term in $F$  should be ignored as it does not affect the equation of motion, so  the leading  symmetry-breaking term is actually cubic  in $\partial_x\varphi$.

In \eqref{eq:oddTerms}, we have introduced the dimensionless phenomenological parameter $\zeta$ that generally can depend on parameters $J/K$, $\tilde{h}$, and $\tilde{\theta}$.   We will not attempt to write down an explicit microscopic expression for $\zeta$ at this stage, for the  following reason:
While one obvious source for those higher-order terms is the higher-order correction in \eqref{eq:cosThetaFull}, it is not the only one. In particular, the current-related part in the continuum Largangian \eqref{eq:Lagr-cont}, which is linear in $\partial_x\vec{n}$, represents only the leading term in the gradient expansion. 
For the ``true'' underlying lattice problem, there will be higher-order terms representing lattice corrections.  
We will come back to this problem later in Section \ref{sec:SLsimulations} when comparing our predictions to the results of spin-lattice  simulations. Those  simulations implement torques at the level of equations of motion via finite differences on an effective lattice, and once the specific lattice implementation is chosen, the corresponding contribution to the Lagrangian can be determined, which will enable us to extract the microscopic expression for $\zeta$.

It is convenient to introduce  the characteristic velocity $c$ and dimensionless spin-transfer velocity $\tilde{v}$,
\begin{equation}
\label{eq:c-vtilde}
c=\sqrt{JK} Sa/\hbar,\qquad \tilde{v}=v/c,
\end{equation}
and pass to dimensionless space  $\xi=x/a$ and time $\tau=ct/a$. In this notation, the Lagrangian \eqref{eq:LQuadratic1} takes the form
\begin{eqnarray}
\label{eq:LQuadratic}
\tilde{L} &=& \frac{J S^2 }{2}
\int d\xi 
\Big\{ 
\frac{1}{2} (\dot{\varphi} + \tilde{v}\varphi')^2  +\tilde{h}\tilde{v}  \frac{\tilde{\zeta}}{3} \varphi'^3 \nonumber\\
&-& (1-\tilde{h}^2)  \Big[
\frac{1}{4} (\varphi'^4 + \varphi''^2) 
- \frac{\tilde{\theta}^2}{2} \varphi'^2 
\Big] 
\Big\},
\end{eqnarray}
where the dot  and the prime denote $\partial_\tau$ and $\partial_\xi$, respectively, and  for the sake of convenience  we have absorbed all the extra factors (except the overall  $ \tilde{h}\tilde{v}$) into the phenomenological coefficient $\tilde{\zeta}=2\zeta (K/J)^{1/2}$. In dimensionless variables, the  condition \eqref{weak-curr-cond1} on the current takes the form $\tilde{v} \ll  (K/J)^{1/2}$.

To account for dissipative effects, we introduce the Rayleigh dissipation function in a standard manner:
\begin{equation} 
\label{eq:RayleighVector}
R =
\int \frac{dx}{a}
\Big\{
\frac{\hbar S \alpha_G}{2} (\partial_t \vec{n})^2 
+ \beta j^{(s)} a^3 (\partial_t \vec{n} \cdot \partial_x \vec{n})
\Big\},
\end{equation}
where ${\alpha _G}$ is the Gilbert damping constant, and $\beta $ is the non-adiabatic spin transfer torque parameter  \cite{kohno2007gauge,TATARA2008microscopic}.
After reduction to the $\varphi$  field and passing to dimensionless variables, one can recast this as
\begin{equation} 
\label{eq:RayleighExpanded}
    \tilde{R} = \sqrt{JK}S^2 (1-\tilde{h}^2)  \int d\xi \left\{ \frac{1}{2} \alpha_G \dot{\varphi }^2 + \beta \tilde{v} \dot{\varphi} \varphi'  \right\},
\end{equation}
where the overall factor is adjusted so that the equation of motion retains the standard form $\delta \tilde{L}/\delta\varphi=\delta \tilde{R}/\delta \dot{\varphi}$.

\section{Effects of the current-induced spin-transfer torque on a uniform  helix} 
\label{sec:helix}

We start with the analysis of the action of current on the helimagnetic system in a single domain state with certain chirality. As will be shown below, this action can be broken down into three effects: homogeneous rotation of the helix, tilting of the magnetization from the easy plane, and renormalization of the helical pitch.  
Throughout this section, we set the magnetic field to zero and deal solely with the effect of spin-transfer torques.

\subsection{Dissipative torque: homogeneous rotation}
\label{subsec:rotation}

It is easy to see that the equation of motion for $\varphi$ admits a solution of the form 
\begin{equation}
\label{eq:Rotation}
\varphi = \tilde{q}(\xi - \tilde{V}\tau)
\end{equation}
which corresponds to the translational shift of the helix along the positive $\hat x$ direction with the constant velocity $\tilde{V}$ 
and is indistinguishable from a homogeneous rotation of the helix with angular velocity $\omega  =  - \tilde{q} \tilde{V}$. 
It is easy to see that all contributions from $\delta \tilde{L}/\delta\varphi$ contain derivatives of $\varphi$ of the second and higher order and thus vanish for the ansatz \eqref{eq:Rotation}, so this solution is effectively governed by the dissipative terms only. The part coming from the dissipation function reduces to $\alpha_G\dot{\varphi}+\beta\tilde{v}\varphi'=0$, which yields the velocity
\begin{equation}
\label{eq:RotationVelocity}
\tilde{V} = (\beta /{\alpha _G}) \tilde{v}.
\end{equation}
 This so-called sliding motion has been recently demonstrated in micromagnetic simulations \cite{xie2024sliding} for a helicoid stabilized by the Dzyaloshinskii-Moriya interaction, and has been investigated experimentally \cite{kimoto2025current} in the helimagnetic MnAu$_2$.
It is a general feature of the action of the non-adiabatic component ($\beta$-term) of the spin-transfer torque in such systems, present due to the constant spatial gradient of magnetic moments.

\subsection{Adiabatic torque: out-of-plane tilt}
\label{subsec:tilt}

Substituting the ansatz \eqref{eq:Rotation} for the rotating helix into  Eq. \eqref{eq:cosThetaFull} and taking into account the result \eqref{eq:RotationVelocity} for the velocity, one can obtain the deviation of the spins from the easy plane in the form
\begin{equation} 
\label{eq:cosThetaBeta}
    \cos \vartheta  =  \frac{\tilde{q}\tilde{v}}{2} \sqrt{\frac{J}{K}}
\left( {1 - \frac{\beta }{{{\alpha _G}}}} \right).
\end{equation}
Helices with opposite chiralities  will have different signs of the out-of-plane canting under the action of a current, which in turn can be probed by the magnetic field perpendicular to the easy plane.  Thus, although the left and right helices are initially degenerate energetically, this degeneracy is lifted if one applies both current and magnetic field together.

As was pointed out in micromagnetic simulations of helimagnetic MnP in \cite{ohe2021chirality}, the choice of chirality in the field decreasing process is inverted when $\beta  > \alpha _G$, which is consistent with \eqref{eq:cosThetaBeta}. 

\subsection{Change of the helical pitch}
\label{subsec:renormalization}

The equation of motion is satisfied for any value of the helical pitch $\tilde{q}$ in the ansatz \eqref{eq:Rotation}. 
The quasi-stationary state of a helical configuration with constant  $\varphi'=\tilde{q}$ can be obtained by minimizing 
 the potential energy density of the Lagrangian \eqref{eq:LQuadratic} with respect to $\tilde{q}$, which yields 
\begin{equation} 
\label{eq:EvenRenorm}
\tilde{q}=\pm\sqrt{ \tilde{\theta}^2+\tilde{v}^2 }.
\end{equation}
Thus, the application of current favors the increase in the helical pitch, making the spiral ``tighter''. For weak currents, the change of the pitch is quadratic in current.  This renormalization is the same for left and right helices, and is insensitive to the direction of the current.

\section{Chirality switching under the combined effect of magnetic field and current} 
\label{sec:ChirSwitch}

When the magnetic field and electric current are applied simultaneously, the symmetry between the left and right chirality is broken. 
This symmetry breaking, encoded in the $\tilde{h}\tilde{v}\varphi'^3$ term of the Lagrangian \eqref{eq:LQuadratic}, occurs because the electric current tilts spins out of the easy plane $(yz)$ into opposite directions for helicoids with different chiralities, so that a magnetic field applied along the hard axis $x$ can distinguish between  the left and right chirality. 
This  provides a mechanism for chirality switching via a combined application of magnetic field and electric current, which has been 
demonstrated in a number of recent experimental works \cite{jiang2020electric,masuda2024room,ohe2021chirality,Yamaguchi+25,Masuda+25,Masuda+26}. 
In this section, we will be interested in conditions required for such switching.

 Similar to Sec.~\ref{subsec:renormalization}, we look at the potential energy density corresponding to the Lagrangian  \eqref{eq:LQuadratic}, now including the field terms. It is convenient to express it in units of $JS^2/2$ and as a function of the first derivative $\psi\equiv\varphi'$:
\begin{equation}
\label{eq:AsymPot}
 \tilde{\mathcal{U}} = (1-\tilde{h}^2)  \Big[
\frac{1}{4} (\psi^4 + \psi'^2) 
- \frac{\tilde{q}_1^2}{2} \psi^2 
\Big] 
-\frac{\tilde{\zeta}}{3}\tilde{h}\tilde{v}   \psi^3 ,
\end{equation}
where we have introduced the notation
\begin{equation}
\label{eq:q1}
\tilde{q}_1^2=\tilde{\theta}^2+\tilde{v}^2/(1-\tilde{h}^2).
\end{equation}

The left and right potential wells are now asymmetric. For definiteness, we assume $\tilde{v}\tilde{h} > 0$, so the well corresponding to the 
``right'' chirality ($\psi>0$) is energetically preferable. The positions of the right and left well minima $\psi_{R,L}$ are given by
\begin{equation}
\label{eq:psiRL}
\psi_{R,L}= \tilde{q}_1 \left( \frac{f}{2}\pm \sqrt{1+\frac{f^2}{4}}  \right),\quad f=\frac{\tilde{\zeta} \tilde{h}\tilde{v}}{\tilde{q}_1 (1-\tilde{h}^2)}.
\end{equation}

If the system was initially in the left potential well ($\psi<0$), its  state becomes metastable.
One might expect that the left well disappears completely for a sufficiently strong asymmetry, pushing the system into the state with the opposite chirality. 
However, it is quite straightforward to check that this cannot happen for any values of the field and current (see Appendix \ref{app:brute-switch}), so such ``brute-force'' chirality switching is not possible.

There is a different switching mechanism, namely via nucleation of the stable phase against the ``false vacuum'' background. 
Let us obtain a rough estimate of the switching conditions for this mechanism.
Assume that we have a domain in energetically unfavorable (metastable) left-chiral state, and we create a pair of chiral domain walls so that between them there is a nucleus of the energetically preferable (stable) right-chiral domain.  Such pairs will be spontaneously created thermally.
Each of the DWs of the form  \eqref{eq:HubertDW} has the characteristic thickness $\ell \sim 1/\tilde{\theta}$, which will be renormalized to $\ell  \sim 1/\tilde{q}_1$ in the presence of the field and current. Thus, for such a pair of kinks to form, they have to be separated by the minimal initial distance of about  $\ell$. The energy cost of creating such a pair 
\begin{equation} 
\label{eq:Epair}
E_{\text{pair}}\simeq 2\tilde{E}_{DW}+   \Delta \tilde{\mathcal{U}} / \tilde{q}_1
\end{equation}
can be estimated as the sum of the penalty for creating the DWs  and the  gain from creating the nucleus of stable phase. 
Here $\Delta \tilde{\mathcal{U}}$ is the energy difference between the left and right wells,
\begin{equation}
\label{eq:Delta}
\Delta \tilde{\mathcal{U}} =\tilde{\mathcal{U}}(\psi_R)-\tilde{\mathcal{U}}(\psi_L)=-\frac{2}{3}\tilde{q}_1^4 (1-\tilde{h}^2) f \left(1+\frac{f^2}{4}\right)^{3/2}.
\end{equation}
The DW energy $\tilde{E}_{DW}$ can be roughly estimated  by variating the potential $\tilde{\mathcal{U}}$ with the symmetry-breaking cubic term ignored (the leading correction due to it is already accounted for in the second term in \eqref{eq:Epair}), essentially following the derivation above that leads to \eqref{eq:ChiralDWEnergy}; expressed in the same units of $JS^2/2$ as $\tilde{\mathcal{U}}$,  this yields
\begin{equation}
\label{eq:EDW}
\tilde{E}_{DW}\sim 2\tilde{q}_1^3 (1-\tilde{h}^2)/3.
\end{equation}
As a rough order-of-magnitude estimate of the switching threshold, one can  look at the point where the total energy cost  $E_{\text{pair}}$ changes sign to negative: this  happens at $f\approx 1.2$.
For a finite system with open edges, one can consider a similar nucleation mechanism that works by creating a single domain wall at the  edge and sweeping it through the system; the corresponding switching condition will have the form similar to  $E_{\text{pair}}<0$, but without the factor $2$ in the first term of \eqref{eq:Epair}, and will be realized for $f\gtrsim 0.8$.  

Our estimate for the switching condition thus reads $f>f_c$, where $f_c$ is some number of the order of unity. To translate this requirement into an explicit domain in the current-field plane, one needs to know the dependence of the phenomenological parameter $\tilde{\zeta}$ on model parameters. We will return to that  later in Section \ref{sec:SLsimulations}.

\section{Chiral domain wall dynamics} 
\label{sec:dwdynamics}

In this section we consider the motion of a single chiral domain wall under the action of applied field and current. 
To that end, we utilize the collective coordinates approach. We introduce the simplest ansatz for the moving DW, based on \eqref{eq:HubertDW}:
\begin{equation}
\label{ansatz-col-coord}
\varphi (\xi ,\tau ) = \pm \ln \cosh \left[ \tilde{q}_1 \big(\xi  - \xi _{DW}(\tau )\big) \right],
\end{equation}
where we have adjusted the DW thickness accounting for the renormalization caused by current and field, and we have a single  collective coordinate 
$\xi _{DW}$ that describes the translational motion of a rigid-shape DW profile. 
Substituting the above ansatz into 
the Lagrangian \eqref{eq:LQuadratic} and the Rayleigh function \eqref{eq:RayleighExpanded}, one obtains
\begin{eqnarray}
\label{LR-col-coord}
\tilde{L}&\simeq &JS^2 \tilde{q}_1^2 \Big\{
\frac{N_x}{4} \dot{\xi}_{DW}^2  \mp \frac{2 \tilde{q}_1\tilde{\zeta}}{3}\tilde{h}\tilde{v} \xi_{DW} \Big\}, \\
\tilde{R}&\simeq& \sqrt{JK} S^2 N_x  \tilde{q}_1^2 (1-\tilde{h}^2) \Big\{
\frac{\alpha_G}{2}\dot{\xi}_{DW}^2 -\beta \tilde{v} \dot{\xi}_{DW} \Big\},  \nonumber
\end{eqnarray}
where $N_x\gg1$ is the dimensionless length of the wire in units of $a$.  Terms with even powers of $\varphi'$ yield contributions proportional to $N_x$ after performing the integration (we neglect corrections of the order of unity). In contrast to that, integrating the term cubic in $\varphi'$ yields contribution proportional to the DW coordinate $\xi_{DW}$ and thus acts as an additional force driving the domain wall.

The equation of motion  has the solution $\xi _{DW}= \tilde{V} \tau$ corresponding to a stationary movement with the constant velocity 
\begin{equation} 
\label{eq:DWVelocity1}
    \tilde{V} = \frac{\beta }{\alpha _G} \tilde{v} \mp \frac{\tilde{\zeta} \tilde{h} \tilde v \tilde{q}_1 (J/K)^{1/2} }{3 \alpha _G N_x (1-\tilde{h}^2)}.
\end{equation}

The first term in \eqref{eq:DWVelocity1} is  the dissipative contribution to the drift velocity found previously in Sec.~\ref{subsec:rotation} for a single helix, see  \eqref{eq:RotationVelocity}. A rough numerical estimate shows that this velocity is not very high:
assuming fully polarized current $P=1$, $S=1$, the lattice constant  $a=0.5 \text{nm}$, and $\beta  \approx \alpha _G$, one gets the relation $ V_\text{diss} \approx 4.0 \cdot  10^{ - 10}  j^{(c)} $, where ${j^{(c)}}$ is the density of the electric current in $\text{A/m}^2$ and $V_\text{diss}$ is in $\text{m/s}$. Therefore, for currents up to $\sim 10^{12}  \text{A/m}^2$ the chiral DW can potentially reach the velocity up to $\sim400 \text{m/s}$.

The second term comes from  the energy difference between two chiral domains, which creates the additional driving force moving the domain wall to expand the favorable domain. This effect, however, is strongly suppressed: from \eqref{LR-col-coord} one can see that both the effective mass of the DW and the damping are macroscopically large (they scale linearly with the longitudinal system size $L_x=N_x a$), while the force acting on the DW remains finite. The resulting contribution to the stationary DW velocity (the second term in  \eqref{eq:DWVelocity1}) is thus suppressed by the factor $1/N_x$ and is relevant only for very small systems with
\begin{equation} 
\label{eq:NxInequality}
    {N_x} \lesssim   \frac{\tilde{h} \tilde{q}_1}{\beta  (1-\tilde{h}^2) }\frac{  \tilde{\zeta} (J/K)^{1/2}}{3}.
\end{equation}
 This contribution can be derived in an alternative way by considering the balance between the energy dissipation during the domain wall motion and the corresponding energy gain due to the energy difference of the chiral domains: equating the energy gain rate $JS^2 \tilde{q}_1^3\tilde{\zeta}\tilde{h}\tilde{v} \tilde{V}_{DW}/3$ to the energy dissipation rate $2\tilde{R}=\sqrt{JK} S^2 N_x  \tilde{q}_1^2 (1-\tilde{h}^2) \alpha_G \tilde{V}_{DW}^2$, one obtains the same expression for the DW velocity $\tilde{V}_{DW}$ as given by the second term in \eqref{eq:DWVelocity1}.

\section{Spin-lattice simulations} 
\label{sec:SLsimulations}

To supplement our theoretical analysis, we perform spin-lattice simulations of a helimagnet, including the effects of both adiabatic and non-adiabatic spin-transfer torque  and  of a magnetic field.  We work with  discretized Landau-Lifshitz equations for unit spin vectors $\vec{n}_i$ living on sites $i$ of the one-dimensional lattice:
\begin{eqnarray}
    \label{eq:MMEquation}
         \frac{d \vec{n}_i}{d t}&=&
\frac{1}{\hbar S} \left( \vec{n}_i \times \vec{h}_{\text{eff}}^i \right) - \alpha_G \left( \vec{n}_i \times \frac{d \vec{n}_i}{d t} \right) \nonumber\\
        &-& \frac{j^{(s)} \mathcal{V}_0}{\hbar S}
        \Big\{
\Delta_x \vec{n}_{i} + \beta 
            \big( \vec{n}_i \times \Delta_x \vec{n}_{i} \big)
        \Big\},
 \end{eqnarray}
where the effective field $\vec{h}_{\text{eff}}^i=-\partial \mathcal{H}_{\text{1d}}/\partial \vec{n}_i $ defined via the 1d lattice Hamiltonian
 \begin{eqnarray}
        \mathcal{H}_{\text{1d}} &=& \sum_i 
        \Big\{ 
            -JS^2(\vec{n}_i \cdot \vec{n}_{i+1}) - J'S^2(\vec{n}_i \cdot \vec{n}_{i+2})  \nonumber \\
            &+& KS^2(\vec{n}_i^x)^2 
             - hS \vec{n}_i^{x} )
        \Big\},    
 \end{eqnarray}
and the effects of spin-transfer torque (the second line in \eqref{eq:MMEquation}) are implemented by discretizing the corresponding continuum expressions via replacing the spatial derivative $\partial_x \vec{n}$ at site $i$ by the symmetric finite difference 
\begin{equation}
\label{eq:DiscreteDerivative}
\Delta_x \vec{n}_{i}=    (\vec{n}_{i+1}- \vec{n}_{i-1}) /(2a).
\end{equation}

For  most simulations, we use the following model parameters: $K = 50 \, \text{meV}$ (typical for rare earth elements \cite{rhyne1967magnetic}), $J = 10 \, \text{meV}$, $J' \approx 2.50955 \, \text{meV}$ (chosen so that  the angle between adjacent spins is $\tilde{\theta} \approx5^{\circ}$, or $\tilde{\theta} \approx 0.087$), $a = 0.5 \, \text{nm}$, and $\alpha_G = 0.05 $, and the system size is $N_x = 200$ spins.  For the sake of simplicity, we set the volume of the unit cell $\mathcal{V}_0=a^3$, the spin value $S = 1$, and assume that the current is fully polarized $j^{(s)}= j^{(c)}\hbar/(2e)$, i.e., $P=1$.  
Whenever deviating parameter values are used for some reason (e.g., when we study the dependence on $N_x$ or $\beta$), it is explicitly indicated in the text.  We solve \eqref{eq:MMEquation} using a simple fourth-order Runge-Kutta integration scheme with a time step $\Delta t = 10^{-3} \, {\rm ps}$. Free boundary conditions are used at the edges.

\subsection{Lattice equations and parameters of the continuum model}

\begin{figure}[tb]
\includegraphics[width=0.85\linewidth]{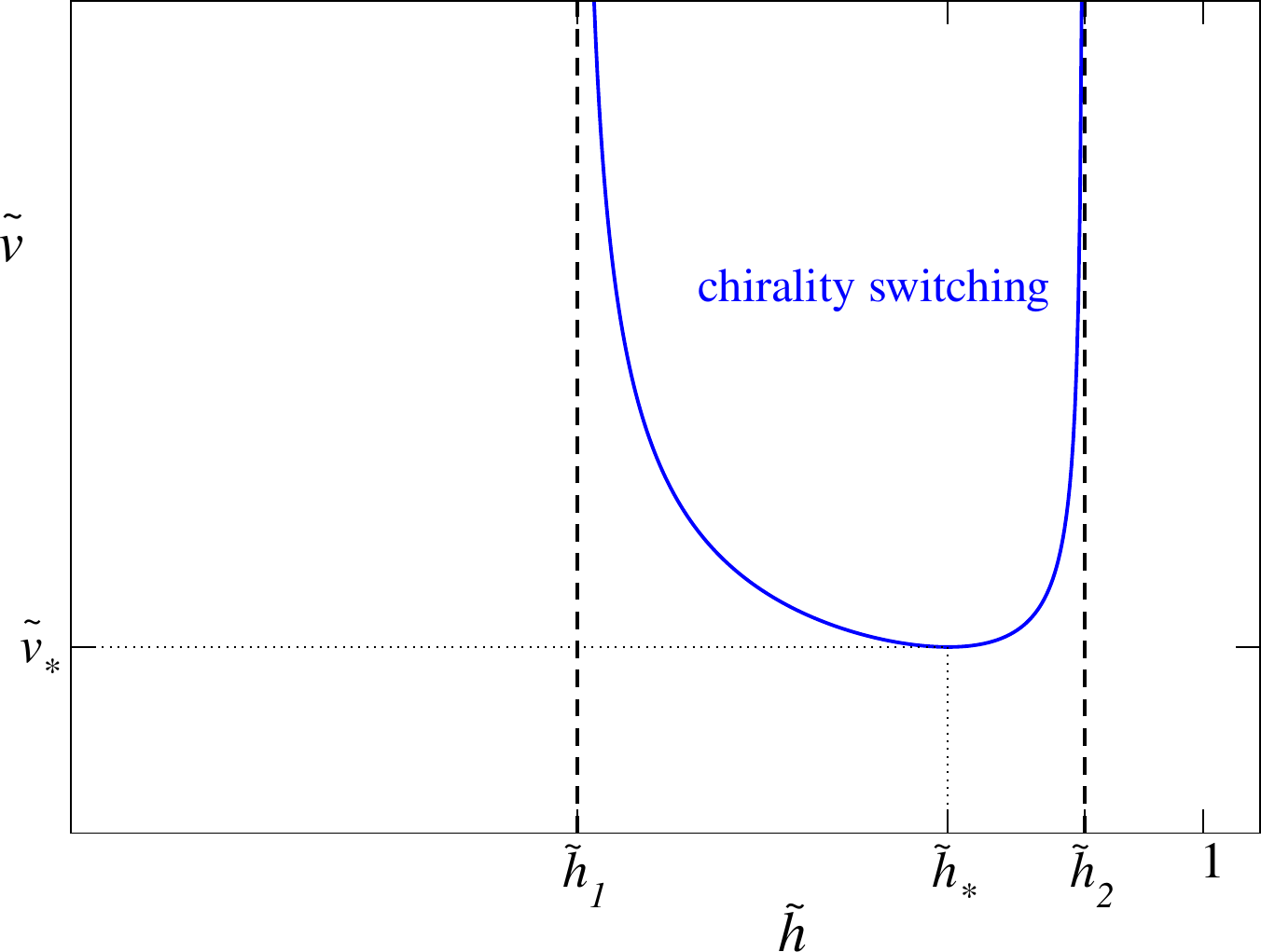}
\caption{
\label{fig:chirswitch}
Schematic view of the region of parameters in the current-field plane where the chirality switching occurs via the domain nucleation mechanism, as given by the condition \eqref{eq:csw1}.
}
\end{figure}

Before we proceed to presenting the simulations results, 
let us look in more detail at the correspondence between the lattice equations  \eqref{eq:MMEquation} and the continuum field description  
used in our theoretical analysis in previous sections. 
We have seen that the chirality switching conditions, as well as the domain wall dynamics, depend crucially on the term in the Lagrangian that is cubic in $\partial_x\varphi$. To be able to compare the simulations results with theory, we in particular need to know how the phenomenological quantity $\tilde{\zeta}$ (the overall factor in front of the cubic term) depends on the model parameters.

In the continuum approach, we have treated the adiabatic part of the spin-transfer torque  within the Lagrangean formalism. It was represented by the term
\begin{equation}
\label{eq:STcont}
L_{st}^{\text{cont}}= -j^{(s)} S_\perp  \int dx \vec{A} \cdot \partial_x \vec{n} 
\end{equation}
in \eqref{eq:Lagr-cont} containing the Dirac monopole vector potential $\vec{A}$ (the cross-section area $S_\perp$ here can be set to $a^2$ for the 1d lattice).  
As we will shortly see, the main contribution of the $(\partial_x\varphi)^3$ type originates from lattice corrections to the above term.
We therefore need to  find out what contribution to the discrete Lagrange function would correspond to the lattice implementation of the adiabatic spin-transfer torque term (the first term in the second line of \eqref{eq:MMEquation}). 
The answer turns out to be fairly simple: the continuum expression \eqref{eq:STcont}  must be replaced by the lattice Berry phase
\begin{equation}
\label{eq:STlat}
L_{st}^{\text{lat}}= - j^{(s)} S_\perp F,\quad F=\sum_i \Omega(\vec{n}_0,\vec{n}_i,\vec{n}_{i+1}),
\end{equation}
where $\Omega(\vec{a},\vec{b},\vec{c})$ is  the signed area of the spherical triangle spanned over the vertices determined by unit vectors  $\vec{a}$, $\vec{b}$, $\vec{c}$  (the oriented solid angle):
\begin{equation} 
\label{eq:Omega}
\tan\frac{\Omega(\vec{a},\vec{b},\vec{c})}{2} = \frac{ (\vec{a},\vec{b},\vec{c})}{1+\vec{a}\cdot\vec{b}+\vec{a}\cdot\vec{c}+\vec{b}\cdot\vec{c}}.
\end{equation}
One can show that
\begin{equation}
\vec{n}_i\times \partial F /\partial \vec{n}_i  =  (\vec{n}_{i+1} - \vec{n}_{i-1})/2,
\end{equation}
which yields precisely the form of the adiabatic spin-transfer torque term used in  \eqref{eq:MMEquation}.  
Note that the equations of motion do not depend on the  arbitrary quantization axis $\vec{n}_0$. 
It should be remarked that this result relies crucially on the symmetric implementation of the derivative \eqref{eq:DiscreteDerivative}. For 
non-symmetric implementations, e.g., the forward difference $(\vec{n}_{i+1}-\vec{n}_i)/a$, it is not possible to represent the adiabatic spin-transfer torque term as a variation/gradient. One may argue, however, that \eqref{eq:STlat} is the proper expression that follows naturally if one attempts to derive the Lagrangian description on a lattice, see Appendix \ref{app:Omega}; consequently, non-symmetric discretizations  of the adiabatic spin-transfer torque term should be avoided in spin-lattice simulations, as they would introduce unphysical contributions.

\begin{figure*}[tb]
\includegraphics[width=0.42\linewidth]{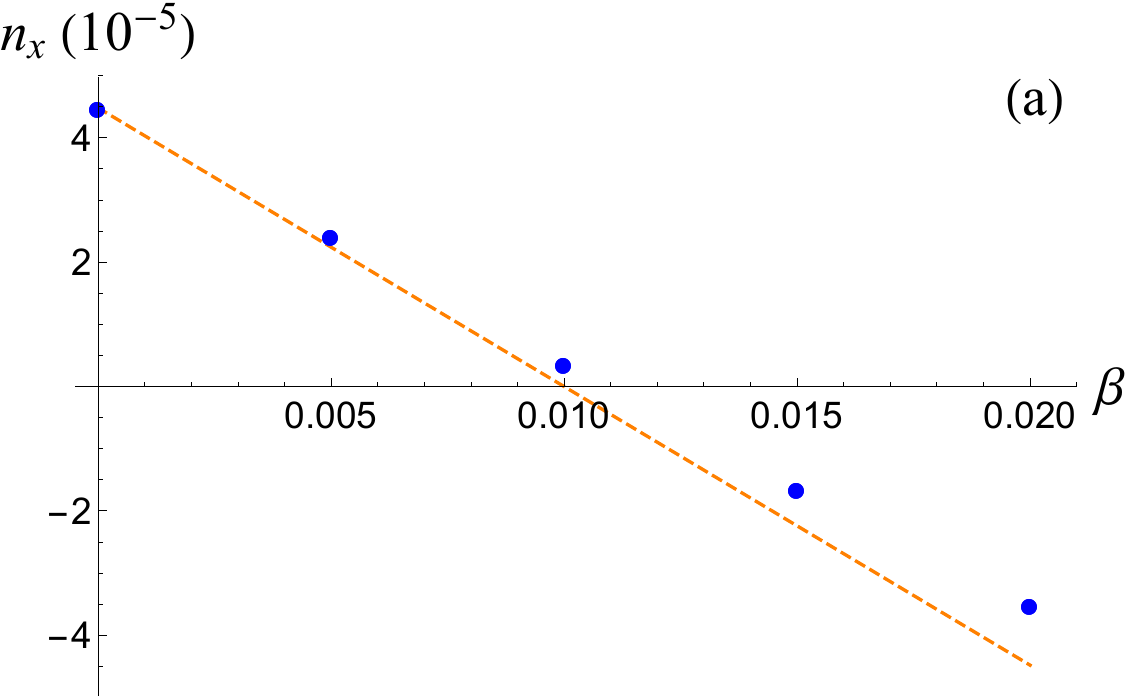}
\includegraphics[width=0.42\linewidth]{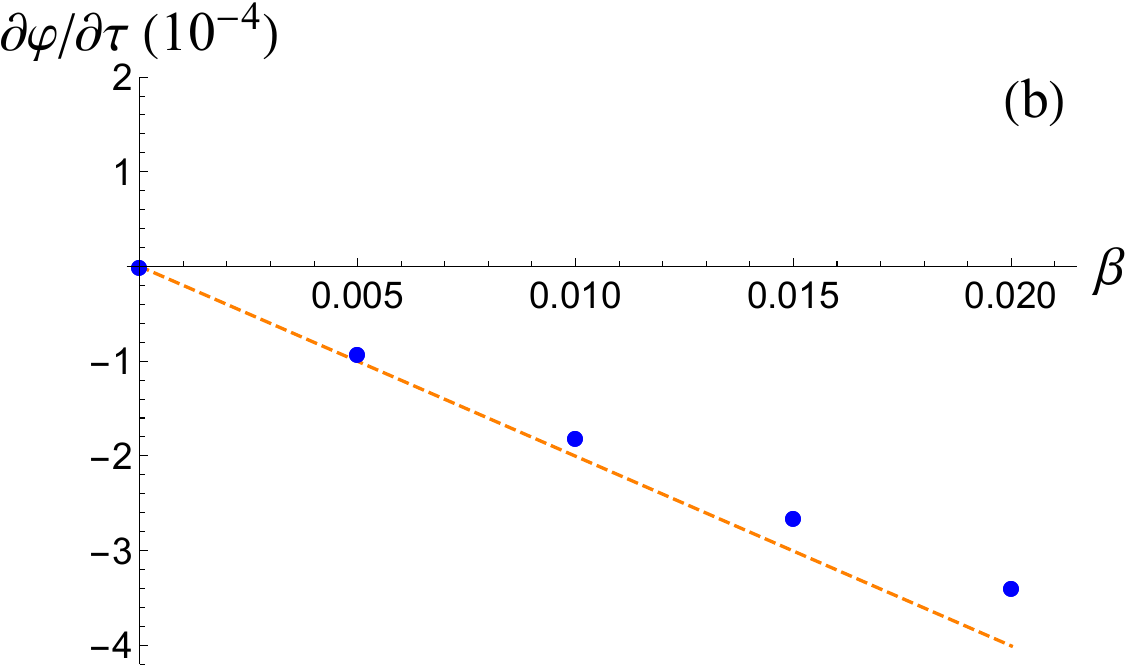}
\caption{
\label{fig:beta}
Results of spin-lattice simulations (circles) compared to analytical predictions from continuum theory (lines)
for the effects of the non-adiabatic spin transfer torque.
 The current  is set  to $j^{(c)} = 10^{11} \, \text{A/m}^2$ ($\tilde{v}\simeq 2.3\cdot 10^{-3}$), the Gilbert damping is chosen as $\alpha_G =0.01$, and the field $\tilde{h}=0$. 
(a) Dependence of the out-of-plane magnetization component $n_x$,  taken at the center of the system, on the non-adiabatic spin transfer torque parameter $\beta$. The line corresponds to \eqref{eq:cosThetaBeta} with $\tilde{q}$ given by \eqref{eq:EvenRenorm}.  (b) Dependence of the angular velocity of the central spin  on $\beta$, the line corresponds to  $\omega  =  - \tilde{q} \tilde{V}$ with $\tilde{V}$ defined in \eqref{eq:RotationVelocity}.
}  
\end{figure*}

Now we can establish the lattice corrections (higher-order terms) to the continuum Lagrangian originating from  \eqref{eq:STlat}. 
We pass to continuum fields, $\vec{n}_i\mapsto \vec{n}(x)$, $\vec{n}_{i+1}\mapsto \vec{n}(x+a)$ and expand the solid angle,  keeping up to the third term in gradients.  It is convenient to fix the Dirac string along the hard axis $\vec{n}_0=\hat{\vec{x}}$, pass to spherical angles $\vartheta$, $\varphi$ for $\vec{n}(x)$, and pass to the dimensionless coordinate $\xi=x/a$. One obtains
\begin{eqnarray}
\label{eq:Omega-cont}
&&\Omega \big(\hat{\vec{x}},\vec{n}(\xi),\vec{n}(\xi+1) \big) =  
(1-\cos\vartheta) \varphi'  \nonumber\\
&&\qquad +\frac{1}{2} \big( (1-\cos\vartheta) \varphi' \big)' \nonumber\\
&& \qquad -\frac{1}{12}\sin^2\vartheta\cos\vartheta (\varphi')^3\\
&&\qquad +\frac{1}{4}\sin\vartheta (\varphi'\vartheta')' +\frac{1}{6}(1-\cos\vartheta) \varphi'''  + \ldots \nonumber
\end{eqnarray} 
The first term in the above expansion is precisely  $\vec{A} \cdot \partial_\xi \vec{n}$. The second-order term is the full derivative and thus should be ignored. Since $\vartheta$ is a slave variable in our theory, see \eqref{eq:cosThetaFull}, terms in the fourth line of \eqref{eq:Omega-cont}  
have effectively higher than the third order in gradients of $\varphi$ and can be omitted as well. 

The leading lattice correction thus comes from the third line of \eqref{eq:Omega-cont}. Including it into the continuum Lagrangian amounts to the replacement $v\partial_x\varphi \mapsto v\partial_x\varphi +(va^2/12) \sin^2\vartheta (\partial_x\varphi)^3$ in \eqref{eq:Lagr-cont1}. 
Since we only need terms of up to the third order in gradients in $\cos\vartheta$, the factor $\sin^2\vartheta$ in the above-mentioned correction can be replaced by the constant $1-\tilde{h}^2$, which results in  the following modification to  \eqref{eq:cosThetaFull}:
\begin{eqnarray}
\label{eq:cosThetaNew}
\cos\vartheta &\simeq &\tilde{h} +\frac{1}{2}\Big( \frac{J}{K}\Big)^{1/2} 
\Big[  \dot{\varphi} + \tilde{v}\varphi' +\frac{\tilde{v}(1-\tilde{h}^2)}{12} \varphi'^3\Big] \nonumber\\
&-&\tilde{h} \frac{J}{4K} \tilde{\theta}^2 \varphi'^2 + \ldots ,
\end{eqnarray}
Substituting   \eqref{eq:cosThetaNew} back into the Lagrangian yields \eqref{eq:LQuadratic} with the following parameter $\tilde{\zeta}$:
\begin{equation}
\label{eq:zeta-micro}
\tilde{\zeta}= \frac{1}{2}\Big( \frac{K}{J}\Big)^{1/2} (1-\tilde{h}^2)  - 3\Big( \frac{J}{K}\Big)^{1/2} \tilde{\theta}^2 + \ldots.
\end{equation}
The second term in the above expression originates from higher-order corrections to $\cos\vartheta$ (cf. the second term in curly brackets in \eqref{eq:cosThetaFull}). Under the adopted assumptions it is much smaller than the first term, unless $\tilde{h}$ gets very close to $1$.  The main contribution to the $(\partial_x\varphi)^3$ term that breaks the chiral symmetry thus comes from lattice corrections to the Berry phase.

With the knowledge of how $\tilde{\zeta}$ depends on the model parameters, one can now translate the chirality switching condition  obtained in Section \ref{sec:ChirSwitch}
\begin{equation}
\label{eq:csw}
\tilde{\zeta} \tilde{h}\tilde{v} > f_c (1-\tilde{h}^2) \sqrt{ \tilde{\theta}^2+\tilde{v}^2/(1-\tilde{h}^2)},\qquad f_c \sim 1,
\end{equation}
into a specific restriction on the values of the current $\tilde{v}$ and field $\tilde{h}$.  
Retaining for simplicity just the leading (first) term in \eqref{eq:zeta-micro}, and substituting it into \eqref{eq:csw}, one obtains
\begin{equation}
\label{eq:csw1}
 \tilde{v} > \kappa\tilde{\theta}\sqrt{ \frac{1-\tilde{h}^2}{\tilde{h}^2 (1-\tilde{h}^2)  -\kappa^2} },\qquad
\kappa=2f_c\sqrt{ \frac{J}{K}}.
\end{equation}

This condition determines the region of parameters in the current-field plane where the chirality switching occurs, as shown schematically in Fig. \ref{fig:chirswitch}.  The switching is possible only for $\kappa<1/2$, i.e., it is realized only if the easy-plane anisotropy is sufficiently high, $K/J>(4f_c)^2$. 
To switch chirality, the magnetic field has to be inside the interval  $[\tilde{h}_1, \tilde{h}_2]$, where $\tilde{h}^2_{1,2}=(1\mp\sqrt{1-4\kappa^2})/2$. 
The minimum current necessary for switching is achieved at the optimal value of the field $\tilde{h}_*=\sqrt{1-\kappa}$ and is given by $\tilde{v}_* = \kappa\tilde{\theta}/\sqrt{1-2\kappa}$. Thus, switching via the nucleation of lower-energy domain requires strong magnetic fields, strong currents, and extremely strong anisotropy.

\subsection{Results of lattice simulations vs continuum theory predictions}

\begin{figure*}[tb]
\includegraphics[width=0.42\linewidth]{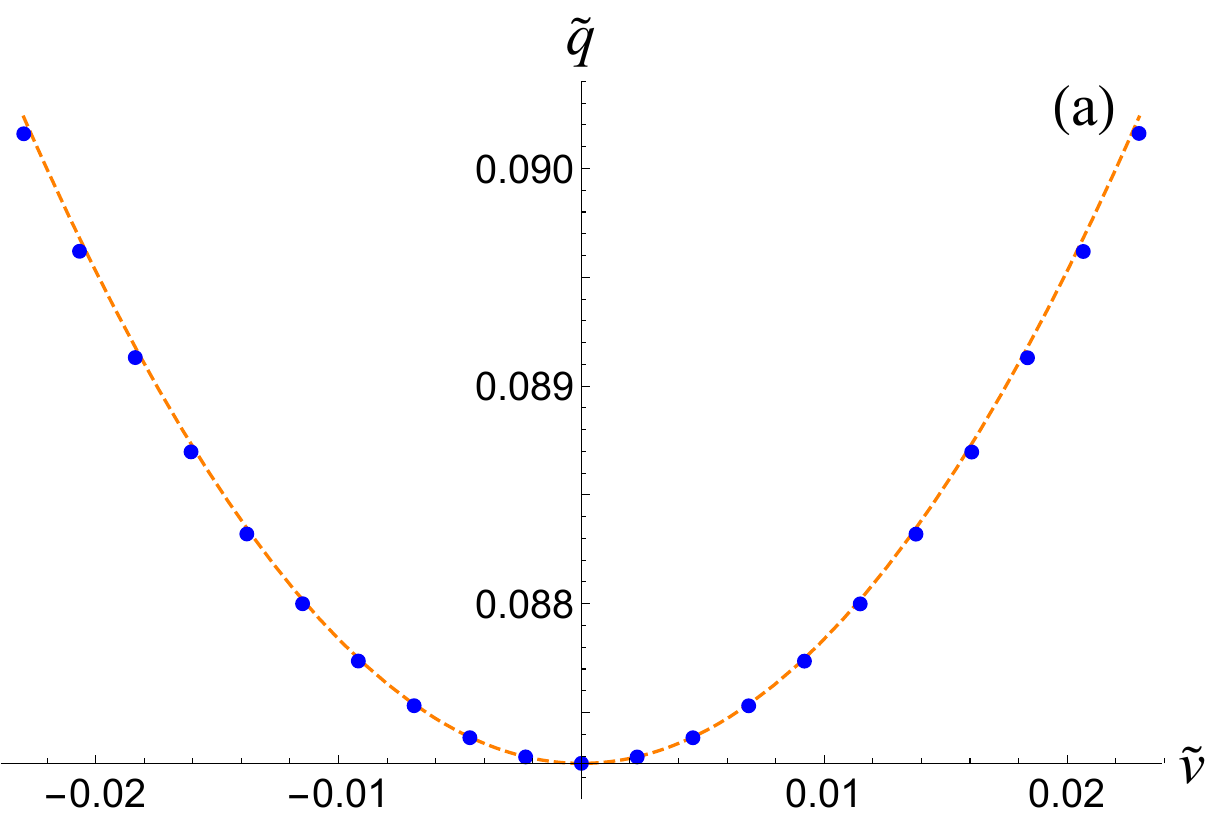}\hspace*{5mm}
\includegraphics[width=0.42\linewidth]{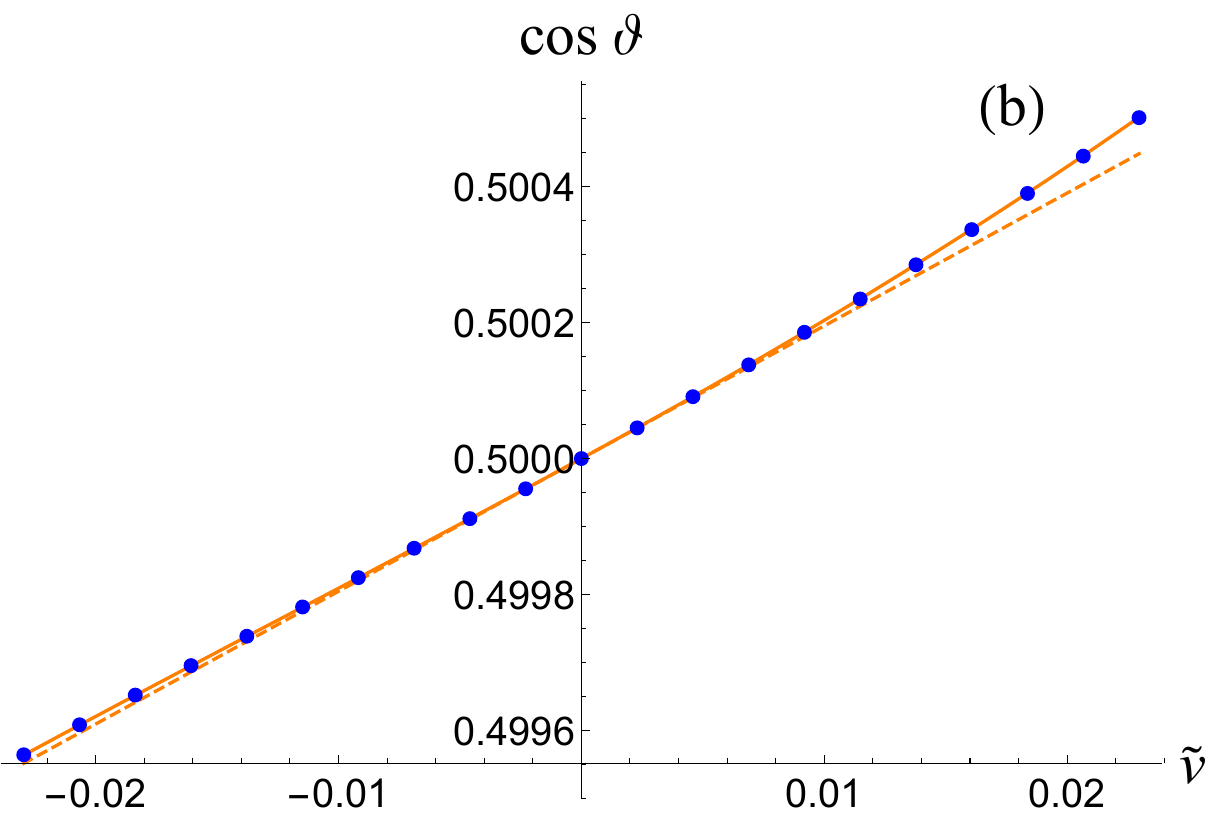}
 \caption{
\label{fig:Mag-and-Pitch}
Results of spin-lattice simulations (circles) compared to analytical predictions from continuum theory (lines) for the effects of adiabatic spin-transfer torque. 
The non-adiabatic spin transfer torque component is switched off, $\beta=0$.
(a) Dependence of the renormalized helical pitch $\tilde{q}$  on the applied current $\tilde{v}$. In the absence of magnetic field $\tilde{h}=0$
 the renormalization is insensitive to the sign of the current. The dashed line corresponds to \eqref{eq:EvenRenorm}.  
(b)  Dependence of the out-of-plane magnetization $\cos\vartheta$ on the applied current $\tilde{v}$, for the fixed value of the magnetic field $\tilde{h}=0.5$. The dashed line corresponds to \eqref{eq:cosThetaNew} with the initial pitch $\varphi'=\tilde{\theta}$, and the solid curve accounts for the renormalized equilibrium value of the pitch $\varphi'=\tilde{q}_1$ according  to \eqref{eq:q1}.
}  
\end{figure*}

We start with checking the continuum theory predictions for the effect of a dissipative spin-transfer torque, in the absence of the external field. We initialize the system in a right-handed helicoidal state, apply the electric current, and read out the out-of-plane magnetization $n_x$ (or $\cos\vartheta$) and the angular rotation frequency  
$\partial_\tau \varphi$ at  the center of the system. The results, shown in Fig.\ \ref{fig:beta}, demonstrate good agreement with the analytical formulas \eqref{eq:cosThetaBeta}, \eqref{eq:Rotation}; in particular, one can see that the  out-of-plane magnetization changes sign at $\beta \approx \alpha_G$, in accordance with \eqref{eq:cosThetaBeta}. 

\begin{figure*}[bt]
\includegraphics[width=0.42\linewidth]{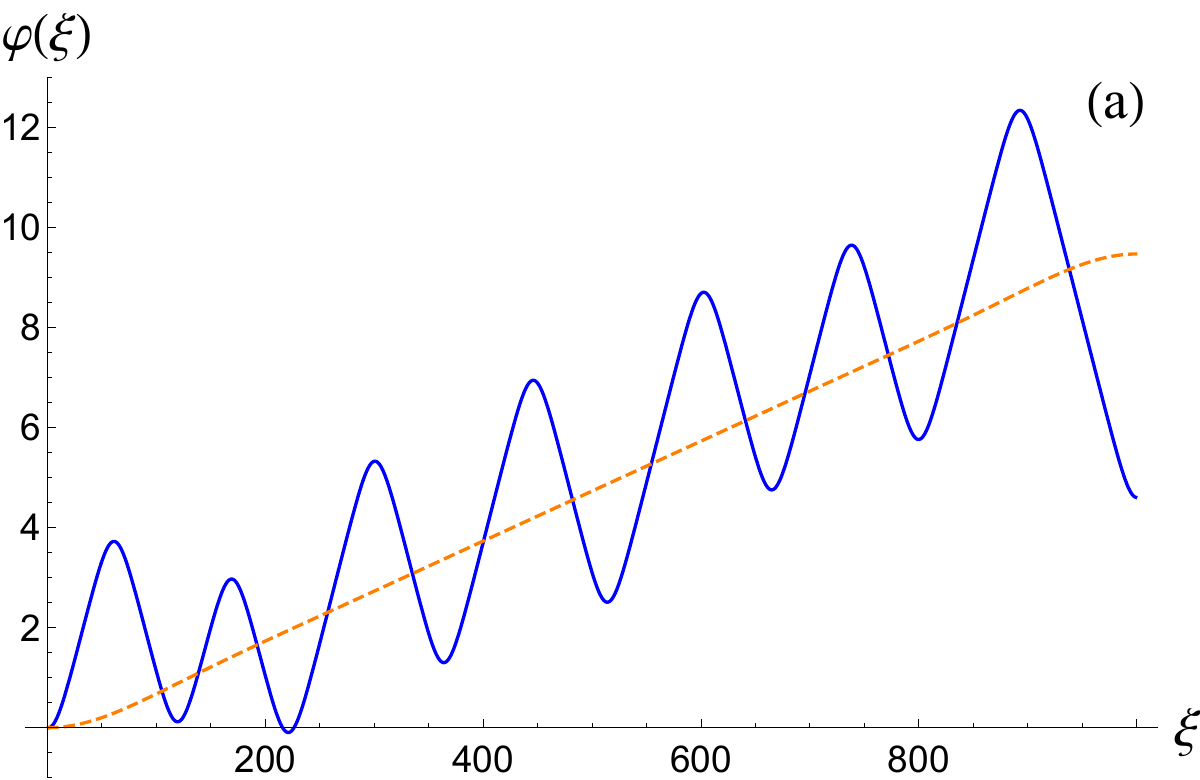}
\includegraphics[width=0.42\linewidth]{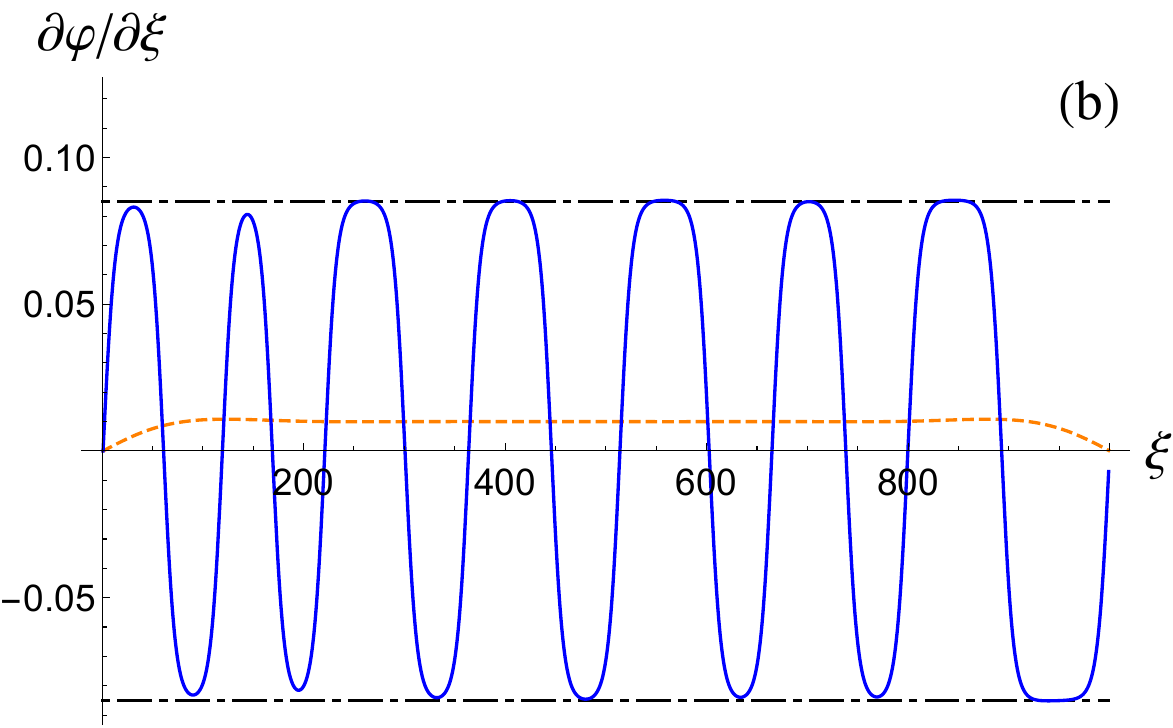}
\caption{ 
\label{fig:phi}
Results of spin-lattice simulations of the system with $N_x=1000$ spins  with zero magnetic field $\tilde{h}=0$.
 A uniform right-handed helix with $\tilde{\theta} =0.01$ has been chosen as the initial state, and the current ${j^{(c)}} = 3.7 \cdot {10^{12}} \ \text{A/m}^2$ (corresponding to $\tilde{v}\approx 0.085$ in dimensionless units) has been applied over the time interval $\Delta t = 100 \ \text{ns}$.  For the sake of clarity, the results are shown as continuous curves interpolating between individual lattice sites (corresponding to integer values of the dimensionless coordinate $\xi  = x/a$).
 (a) The resulting evolution of $\varphi (\xi )$.  The dashed line corresponds to the initial single-domain state, and the solid curve shows the final state  consisting of a sequence of helical domains with alternating chiralities and increased pitch.
(b) The same evolution for  the slope $\varphi' (\xi)$. The initial state corresponds to the dashed line, the solid line shows the final state, and dash-dotted lines indicate the renormalized values of the pitch $\pm \sqrt{\tilde{\theta}^2 +\tilde{v}^2}$.
}
     
\end{figure*}

There is a moderate quantitative discrepancy which we attribute to the fact that the system becomes slightly inhomogeneous under the action of a dissipative torque (e.g., the helical pitch $\varphi'$ varies along the chain). It is the sign of a nucleation of a different dynamical mode, which appears for larger currents, or, more generally, larger values of $\beta$-torque: instead of a (quasi-)homogeneous rotation, the helix 'unwinds' into a periodic multi-domain state from the boundary (if the length of the sample is sufficient, $N_x\gg 1 / \tilde{\theta}$), then the whole magnetic profile moves across the system in the form of a traveling wave with the velocity $\tilde{V}$ \eqref{eq:RotationVelocity}. The form of the solution is similar to that shown in Fig.\ \ref{fig:phi}, although in this case it is not stationary and its origin is not the renormalization of the pitch, but the action of a strong dissipative torque. Note that the initial chiral state is being rewritten in such a process, and the average chirality now becomes zero (if we do not consider the finiteness of the sample).

We further set $\beta=0$ and check the continuum theory prediction for the effects of adiabatic spin-transfer torque. 
Without magnetic field ($\tilde{h}=0$) the obtained helical pitch increases quadratically with the applied current in a good agreement with 
 \eqref{eq:EvenRenorm}, as shown in  Fig.\ \ref{fig:Mag-and-Pitch}a. 
In the absence of the field, the renormalization of the pitch does not depend on the direction of the current, and is the same for right and left helicoids.
The dependence of the out-of-plane magnetization on the applied current agrees very well with the prediction \eqref{eq:cosThetaNew} as demonstrated in Fig.\ \ref{fig:Mag-and-Pitch}b.

It should be remarked that the adjustment of the ground state to the renormalized pitch is not trivial, as the equation of motion does not contain the ``returning force'' which would simply push the system toward the potential minimum, and the overall dynamics of the magnetization is strongly restricted due to the easy-plane confinement. In simulations, we observe that the actual perturbation  starts from the boundaries (which are naturally inhomogeneous due to the absence of neighbors on one side) and then propagates along the system.
For weak currents $(\tilde{v}\ll\tilde{\theta})$ the propagating disturbance smoothly changes the slope $\varphi'$, thus leading to a uniform renormalization of the pitch. However, for sufficiently strong applied currents, when the renormalized $\tilde{q}$ is a few times larger than the initial value $\tilde{q}_0 = \tilde \theta $, we observe a different mechanism: the initially uniform (single-domain) helical state evolves by breaking into multiple domains with opposite chiralities, adjusted to the new pitch  $\pm\tilde{q}$ and separated by chiral domain walls, as shown in Fig. \ref{fig:phi}. The same behavior can be reproduced by numericallly solving the continuum equation of motion determined by the Lagrangian \eqref{eq:LQuadratic}.

\begin{figure}[tb]
\centering
\includegraphics[width=0.84\linewidth]{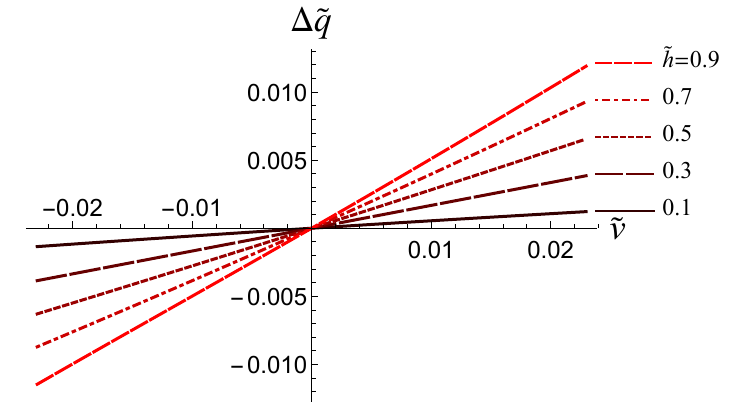}
\caption{
\label{fig:dqvsjh}
The  shift $\Delta \tilde{q}$ of the equilibrium value of the pitch $\psi_R$ under the action of the electric current $\tilde{v}$, for several values of the applied magnetic field $\tilde{h}$. Only the part of the shift that is odd in $\tilde{v}$ is shown (the even component is subtracted).
}
\end{figure}

In the next step, we test the continuum theory predictions on chiral symmetry breaking when the magnetic field and electric current are applied simultaneously.
According to \eqref{eq:psiRL},  the renormalized helical pitch should get a contribution that is odd both in current $\tilde{v}$ and field $\tilde{h}$.
To extract this odd contribution, we proceed as follows.
In \eqref{eq:psiRL}, we keep only the linear term in $f$ (which is justified for small $\tilde{v}$), and substitute the microscopic expression for $\tilde{\zeta}$ keeping only the leading (first) term in  \eqref{eq:zeta-micro}.  We thus obtain the following expression against which we fit the numerically extracted equilibrium pitch for the right-handed helix:
\begin{equation}
\psi_R \approx \sqrt{\tilde{\theta}^2+C_1\frac{\tilde{v}^2}{1-\tilde{h}^2}} +  \frac{C_2}{4} \sqrt{\frac{K}{J}}  \tilde{h}\tilde{v},
\end{equation}
where we have introduced two fitting coefficients $C_{1,2}$ to be able to filter out the even part with better accuracy.  
The fit across a wide range of applied fields ($\tilde{h}$ varying from $0.1$ to $0.9$) yields $C_1\simeq 1.016(13)$ and $C_2\simeq 1.0019(15)$ in a very good agreement with the theoretical prediction  $C_{1,2}=1$. 
Fig.\ \ref{fig:dqvsjh} shows the odd part of the dependence of the renormalized pitch on the current, for different values of the field $\tilde{h}$, demonstrating the predicted linearity in $\tilde{v}$ and  $\tilde{h}$.

\begin{figure}[tb]
\centering
\includegraphics[width=0.84\linewidth]{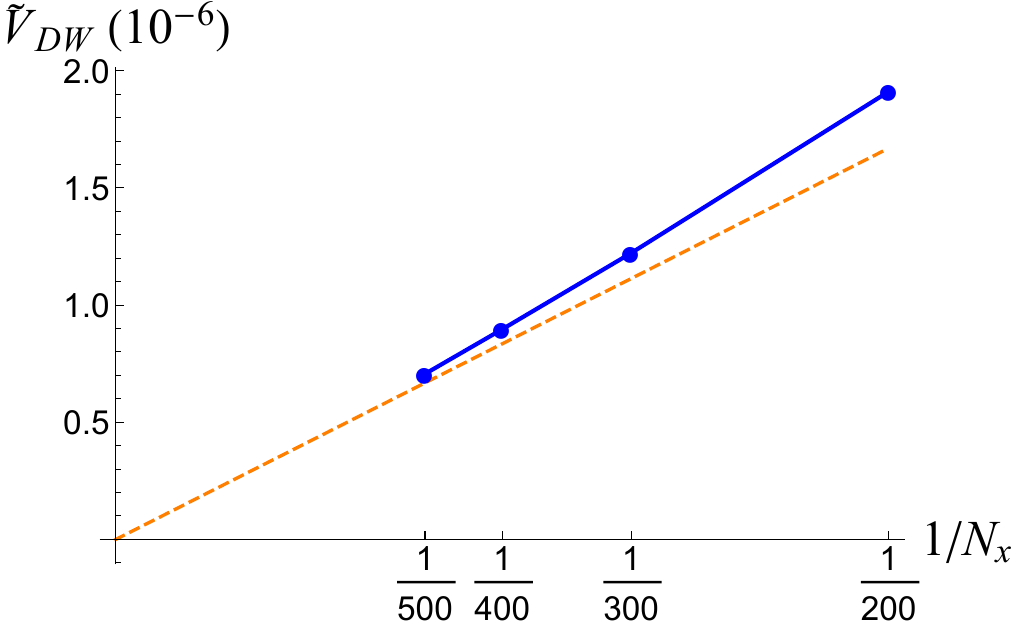}
\caption{
\label{fig:vdwvsovernx}
Dependence of the domain wall velocity $\tilde{V}_{DW}$ on the longitudinal size of the system $N_x$. The dashed line corresponds to the formula \eqref{eq:DWVelocityCubic}, the solid line connecting the dots is a guide to the eye.
}
\end{figure}

Finally, we check the predictions of the continuum theory on the contribution to the domain wall velocity from the term that breaks the chiral symmetry.
It is represented by the second term in \eqref{eq:DWVelocity1}, which after substituting $\tilde{\zeta}$ from  \eqref{eq:zeta-micro} and $\tilde{q}_1$ from \eqref{eq:q1}, yields
 \begin{equation}
\label{eq:DWVelocityCubic}
\tilde{V}_{DW}=\frac{\tilde{h}\tilde{v} }{6\alpha_G N_x}\sqrt{\tilde{\theta}^2+\frac{\tilde{v}^2}{1-\tilde{h}^2}} .
\end{equation}
In simulations, we initialize systems of various sizes from $N_x=200$ up to $N_x=500$ with a chiral domain wall \eqref{eq:HubertDW} in the center, let the configuration relax,   then turn on the current $j^{(c)}=-10^{11}\,\text{A/m}^2$ ($\tilde{v}\simeq 2.3 \cdot 10^{-3}$) and the field $\tilde{h}=0.5$,  track the DW motion until the velocity reaches stationary regime, and then extract the velocity value over the interval near the middle of the system (to avoid the effects of interaction with the boundaries). Fig. \ref{fig:vdwvsovernx} shows the resulting dependence of the stationary velocity on the system size, which agrees with the theoretical prediction \eqref{eq:DWVelocityCubic} in the limit of large $N_x$.

\section{Summary}
\label{sec:conclusion}

We have developed theoretical description of a helimagnet in the planar limit (strong easy-plane anisotropy) under the action of  electric current and external magnetic field. The theory is based on the continuum description valid in the vicinity of the Lifshitz point where the period of the magnetic helix is large compared to the lattice constant. 
It is shown that a chiral domain wall, connecting helical domains of opposite chirality, can be moved by the current via the
dissipative (non-adiabatic) component of the spin-transfer torque, while the adiabatic part of the torque renormalizes the equilibrium pitch of the helix. 
It is demonstrated that the combined effect of magnetic field and current breaks the symmetry between the left-handed and right-handed helical states, providing a  mechanism for chirality switching;  it is further shown that although the current-induced out-of-plane spin canting is linear in the helical wave number $q$ (``pitch'' of the spiral), the leading symmetry-breaking term in the effective potential energy is cubic  in $q$. 
We check those analytical predictions against the results of spin-lattice simulations and find a satisfactory agreement.

It should be remarked that  our theory is not directly applicable to the recent experiments \cite{jiang2020electric,masuda2024room,ohe2021chirality,Yamaguchi+25,Masuda+25,Masuda+26} as they are performed on materials with weak anisotropy. In the planar regime considered in the present work, chirality switching via nucleation of  an energetically favorable domain is hard to realize as it requires very strong easy-plane anisotropy as well as prohibitively strong magnetic fields. Nevertheless, the present work captures the essential physical principle behind the chirality switching, namely, chiral symmetry breaking by a simultaneous application of current and field, and we believe that the insight gained here will prove useful in the future extension of the theory to the regime of weak anisotropy.

\begin{acknowledgments}
We are grateful to Yaroslaw Bazaliy, Revaz Ramazashvili, and Igor Gerasimchuk for valuable comments.  
R.T.  was supported by the Project No. 3F-2026 of  the Department of Targeted Training of Taras Shevchenko National University of Kyiv at the National Academy of Sciences of Ukraine, and  by the Scholarship for Young Scientists of the National Academy of Sciences of Ukraine. 
\end{acknowledgments}

%
\appendix
%
\section{Impossibility of the ``brute force'' chirality switching}
\label{app:brute-switch}

We would like to find out whether it is possible to make the left well of the asymmetric two-well potential \eqref{eq:AsymPot} disappear completely  by applying sufficiently strong magnetic field and electric current. The condition for this to happen can be formulated as follows: find the extrema $\psi_\pm$ of the first derivative $U'=dU/d\psi$
(we ignore  the term with $\psi'^2$ and so consider only homogeneous helical configurations).

The left extremum $\psi_-$ is a maximum. As long as the value of $U'(\psi_-)$ at this  maximum is positive, the left well exists (i.e., $U'(\psi)$ has a zero at some point $\psi_L<\psi_-$). Thus, the left well disappears if $U'(\psi_-)\leq 0$. To check this, it is convenient to reduce the number of free parameters in  \eqref{eq:AsymPot}. We divide the potential by $\tilde{q}_1(1-\tilde{h}^2)$, change the independent variable to $x=\psi/\tilde{q}_1$, and write the rescaled potential as
$u(x)=\frac{1}{4}x^4 - \frac{1}{2}x^2 -\frac{1}{3}fx^3$, 
where  $\tilde{q}_1$  and  $f$ are defined in  \eqref{eq:q1} and \eqref{eq:psiRL}, respectively.
The extrema of $u'(x)$ are given by $x_\pm=(f \pm\sqrt{f^2+3})/3$, and the condition $u'(x_-)\leq 0$  takes the form
\begin{equation}
\label{eq:cond-brute}
 f(\sqrt{f^2+3}-f) \geq 6
\end{equation}
which cannot be satisfied for any value of $f$. Thus, within our approach,  though the combined application of the magnetic field and electric current can make the ``unfavorable'' chiral state metastable, it never becomes unstable, so the chirality switching cannot work this way.

\section{Lattice Lagrangian for the adiabatic spin-transfer torque}
\label{app:Omega}

It is well-known that electrons interacting with an inhomogeneous magnetization texture experience a fictitious “gauge field” that couples to spin current \cite{ShraimanSiggia88,Bazaliy98,kohno2007gauge}. This formalism is usually used directly in the continuum theory framework, and only the leading (linear) term in the texture gradients is kept in the expression for the gauge field.
The goal of this Appendix is to give some hand-waving derivation of the corresponding lattice expressions that would allow to take into account higher-order corrections.

As the starting point, consider a tight-binding 1d lattice model for conduction electrons described by the Hamiltonian
\begin{eqnarray}
\label{tight}
&& H = H_\text{hop} + H_{sd},\qquad H_\text{hop} = -t\sum_{l}(\psi_l^\dagger \psi_{l+1}^{\vphantom{\dagger}} +\text{h.c.})\nonumber\\
&&  H_{sd}= -J_{sd}\sum_l \psi_l^\dagger (\vec{\sigma}\cdot\vec{S}_l) \psi_l^{\vphantom{\dagger}},
\end{eqnarray}
where $l$ numbers lattice sites, $\psi_l=(c^\uparrow_l,c^\downarrow_l)^T$ is the two-component spinor  describing the conduction electrons (in the frame with the quantization axis $\vec{n}_0$), $t$ is the hopping amplitude between nearest neighbor sites, and $J_{sd}$ is the exchange coupling to localized spins $\vec{S_l}=S\vec{n}_l$, where  $\vec{n}_l$ are classical unit vectors.
The Lagrangian corresponding to this tight-binding model can be written as
\begin{equation}
\label{tight-L}
L=i(\hbar/2)\sum_l \left( \psi_l^\dagger \partial_t \psi_{l}^{\vphantom{\dagger}} - \partial_t \psi_l^\dagger \psi_{l}^{\vphantom{\dagger}}\right) -H .
\end{equation}

Performing local unitary transformation $\psi_l =U(\vec{n})\chi_l$ with 
\begin{equation}
\label{unitary}
U(\vec{n}_l)=\vec{\sigma}_l\cdot\vec{e}_l,\quad 
\vec{e}=\frac{ \vec{n}_0+\vec{n}}{[2(1+\vec{n}\cdot\vec{n}_0)]^{1/2}},
\end{equation}
which describes a $180^{\circ}$ rotation about the direction $\vec{e}$ that bisects the angle between $\vec{n}_l$ and $\vec{n}_0$,
we rotate the quantization axis at each lattice site to $\vec{n}_l$, so that the interaction term 
$H_{sd} =S\sum_l \chi_l^\dagger (\vec{\sigma}\cdot \vec{n}_0) \chi_l$ is diagonalized in this twisted frame. This twist modifies hopping as follows:
\begin{eqnarray}
\label{hopping}
&& H_\text{hop}  =  -t \sum_l  \left\{ \chi_l^\dagger \left[  a_l \mathds{1} +i (\vec{b}_l \cdot \vec{\sigma}) \right] \chi_{l+1} +\text{h.c.}\right\},\nonumber\\
&& a_l= (\vec{e}_l\cdot\vec{e}_{l+1}),\qquad \vec{b}_l= (\vec{e}_l\times\vec{e}_{l+1})
\end{eqnarray}
In the adiabatic picture, the spin of electrons follows the direction of localized spins, so in the rotated frame all electron spins are polarized along the axis $\vec{n}_0$. Thus, implying that the average is eventually taken over the electron operators, one can replace $\vec{b}_l$ above by its projection onto this axis, $\vec{n}_0 (\vec{b}_l\cdot \vec{n}_0)$. This allows one to recast  $H_\text{hop}$ as
\begin{eqnarray}
\label{hop2}
&& H_\text{hop}  \mapsto  -  \sum_l t_l  
\left\{  
\chi_l^\dagger e^{i  (\vec{\sigma}\cdot\vec{n}_0) \Omega(\vec{n_0},\vec{n}_l,\vec{n}_{l+1})/2} \chi_{l+1} +\text{h.c.}
\right\}, \nonumber\\
&& t_l=t \left( \frac{1+\vec{n}_l\cdot\vec{n}_{l+1}}{2}\right)^{1/2},
\end{eqnarray}
where $\Omega$ is the oriented solid angle defined in  \eqref{eq:Omega}. Now, the derivative of $H_\text{hop}$ over $\vec{n}_l$ takes the form
\begin{eqnarray}
\label{derhop}
\frac{\partial H_\text{hop}}{\partial \vec{n}_l} &=& \left(\vec{I}^{(s)}_l \cdot\vec{n}_0\right) \frac{\partial \Omega(\vec{n_0},\vec{n}_l,\vec{n}_{l+1})}{\partial \vec{n}_l} \nonumber\\
&+& 
\left(\vec{I}^{(s)}_{l-1} \cdot\vec{n}_0\right) \frac{\partial \Omega(\vec{n_0},\vec{n}_{l-1},\vec{n}_{l})}{\partial \vec{n}_l}, 
\end{eqnarray}
where
\begin{equation}
\vec{I}^{(s)}_l = -\frac{it_l}{2}\left\{  \chi_l^\dagger  \vec{\sigma} e^{i  (\vec{\sigma}\cdot\vec{n}_0) \Omega(\vec{n_0},\vec{n}_l,\vec{n}_{l+1})/2}   \chi_{l+1} - \text{h.c.} \right\}
\end{equation}
is the spin current flowing from site $l$ to site $l+1$, as one can recognize from the time evolution of the spin density at site $l$:
\begin{equation}
\frac{\partial}{\partial t} \left\{ \chi_l^\dagger \vec{\sigma} \chi_{l} \right\} = \left[ \chi_l^\dagger \vec{\sigma} \chi_{l}, H_\text{hop}\right] = \vec{I}^{(s)}_{l-1} - \vec{I}^{(s)}_l .
\end{equation}
Assuming that the average of the spin current is homogeneous, $\langle \vec{I}^{(s)}_l \rangle =I^{(s)}\vec{n}_0$, one can see that  \eqref{derhop} effectively corresponds to the following expression for $H_\text{hop}$ averaged over electrons:
\begin{equation}
\label{eq:Omega-derived}
\langle H_\text{hop} \rangle = I^{(s)} \sum_l \Omega(\vec{n_0},\vec{n}_l,\vec{n}_{l+1}),
\end{equation}
which is equivalent to \eqref{eq:STlat}. It should be remarked that the simple form of the  result  \eqref{eq:Omega-derived} is connected to the simplicity of the initial model \eqref{tight} with only nearest-neighbor hopping, and can be easily generalized to the case when longer-range hopping is present.

\bibliography{refs-curr-heli}
\end{document}